\documentclass[proof]{WileyASNA-v1}

\articletype{Original Article}%

\received{8 July 2022}
\revised{XX July 2022}
\accepted{YY July 2022}

\def\Msun{\ifmmode {\rm M}_{\odot} \else M$_{\odot}$\fi}

\def\nvii{N\,{\sc vii}}

\def\oviiviii{O\,{\sc vii-viii}}

\def\neixx{Ne\,{\sc ix-x}}
\def\fexxv{Fe\,{\sc xxv}}
\def\chandra{{\it Chandra}}
\def\xmm{{\it XMM-Newton}}

\def\extp{{\it eXTP}}
\def\xrism{{\it XRISM}}
\def\athena{{\it ATHENA}}

\begin{document}

\title{Winds in ultraluminous X-ray sources: new challenges}

\author[1]{C. Pinto*}

\author[2]{P. Kosec}


\authormark{C. Pinto and P. Kosec} 


\address[1]{\orgdiv{INAF}, \orgname{IASF}, \orgaddress{Palermo, \country{Italy}}}

\address[2]{\orgdiv{MIT}, \orgname{Kavli Institute for Astrophysics and Space Research}, \orgaddress{Cambridge \state{Ma}, \country{USA}}}

\corres{*C. Pinto, INAF - IASF Palermo, Via U. La Malfa 153, I-90146 Palermo, Italy \email{ciro.pinto@inaf.it}}


\abstract{Ultraluminous X-ray sources (ULXs) are extreme X-ray binaries \textcolor{black}{shining above $10^{39}$ erg/s, in most cases as a consequence of super-Eddington accretion onto neutron stars and stellar-mass black holes accreting above their Eddington limit}. This was understood after the discovery of coherent pulsations, cyclotron lines and powerful winds. The latter was possible thanks to the high-resolution X-ray spectrometers aboard \textit{XMM-Newton}. ULX winds carry a huge amount of power owing to their relativistic speeds (0.1-0.3\,$c$) and are able to significantly affect the surrounding medium, likely producing the observed 100\,pc ULX superbubbles, and limit the amount of matter that can reach the central accretor. The study of ULX winds is therefore quintessential to understand 1) how much and how fast can matter be accreted by compact objects and 2) how strong is their feedback onto the surrounding medium. This is also relevant to understand supermassive black holes growth. Here we provide an overview on this phenomenology, highlight some recent, exciting results and show how future missions such as \textit{XRISM}, \textit{eXTP} and {\athena} will improve our understanding.}

\keywords{X-ray binaries, accretion, accretion discs, black hole physics, stars: winds, outflows}

\jnlcitation{\cname{%
\author{C. Pinto}
and
\author{P. Kosec}} (\cyear{2022}), 
\ctitle{Winds in ultraluminous X-ray sources: new challenges}, \cjournal{Astron. Nachr. / AN.}, \cvol{2022;00:1--6}.}


\maketitle


\section{Introduction}\label{sec:intro}

Ultraluminous X-ray sources (ULXs) are the most extreme among X-ray binaries (XRBs) with 
X-ray luminosities in excess of the Eddington limit for a standard {10\,\Msun} black hole (BH), or $10^{39}$\,erg/s. 
Despite being discovered already 40 years ago with the \textit{Einstein} observatory,
it was only after the launch of powerful telescopes such as
\textit{Chandra}, \textit{XMM-Newton} and \textit{NUSTAR} that we began to understand their nature.

In the early 2000s, ULX extreme luminosities were thought to be mainly produced by 
intermediate-mass $(10^{2-4} M_{\odot})$ black holes (IMBHs)
accreting well below their Eddington limit (e.g., \citealt{Kaaret2001}).
Two alternative scenarios invoked either extreme geometrical beaming in 
Eddington-limited stellar-mass
black holes (BHs, e.g., \citealt{King2001})
or super-Eddington accretion onto compact objects like BHs and neutron stars (NS), 
\textcolor{black}{a common interpretation for the Galactic microquasar SS 433 UV emission} (e.g., \citealt{Begelman2006}).
Both the IMBH and super-Eddington scenarios would be important to understand how it was possible
to find fully-formed supermassive black holes (SMBHs) at high redshifts
when the universe was young (e.g., \citealt{Fan2003, Volonteri2003}).

Dedicated observations with \textit{XMM-Newton} and, later on, \textit{NUSTAR} have shown that 
ULXs are characterised by very soft spectra with a strong curvature below 10 keV,
which would rule out highly sub-Eddington accretion and, therefore, IMBHs as compact objects
(e.g., \citealt{Gladstone2009,Bachetti2013}). 
The so-called \textit{ultraluminous state} is classified according to three main regimes depending
on the spectral slope, $\Gamma$: 
soft (SUL, $\Gamma>2$) or hard (HUL, $\Gamma<2$) ultraluminous. In the latter, 
if the X-ray spectrum has a single peak and a blackbody-like shape, it is called broadened disc 
regime (BD, \citealt{Sutton2013}). 

The first masses inferred dynamically from optical data revealed two ULXs 
to be powered by compact objects with masses in the stellar-mass range 
(e.g., \citealt{Liu2013,Motch2014}). 
This research field was significantly shaken after the first discoveries of pulsations in high-quality
time series of ULXs (e.g., \citealt{Bachetti2014,Fuerst2016,Israel2017a}) 
\textcolor{black}{and cyclotron lines in a few ULX spectra (\citealt{Brightman2018,Walton2018b}).}
This showed that in many ULXs the compact object is actually a magnetised
neutron star rather than a black hole. 

Another ground-breaking discovery was the detection of ULX winds in the form of resolved 
rest-frame emission and blueshifted (0.1-0.3\,$c$) absorption lines in deep ($>$100\,ks) high-resolution X-ray 
spectra taken with the Reflection Grating Spectrometers (RGS) aboard \textit{XMM-Newton}
(e.g, \citealt{Pinto2016nature,Pinto2017,Kosec2018a,Kosec2018b}).
These detections resolved and identified the nature of the spectral features previously spotted
in low-resolution CCD spectra (e.g, \citealt{Stobbart2006}), which were attributed
to the ULX rather than the host galaxy owing to their shape and variability
(e.g, \citealt{Sutton2015,Middleton2015b}).
\textit{Chandra} gratings confirmed similar outflows in a Galactic transient ULX 
(\textcolor{black}{Swift J0243.6+6124,} \citealt{vdEijnden2019}). Further CCD work confirmed the presence of blueshifted absorption
features \citep{Walton2016a,Wang2019}. 
These discoveries confirmed a major prediction of theoretical simulations of super-Eddington 
accretion discs in which radiation pressure is expected to launch mildly-relativistic winds
(e.g., \citealt{Kobayashi2018}).

Despite important discoveries and progress in the last decade, 
several questions remained unanswered. \textit{Are the spectral transitions 
triggered by orbital variations in the accretion rate or by stochastic variations
in the wind? What are the disc-wind geometry and thermal structure?
Does the wind regulate the growth of the compact object and inflate the interstellar cavities?}
To solve these issues a combination of accurate plasma modelling and
spectral-timing studies is required.


\section{Winds: recent developments}\label{sec:developments}

\begin{figure}[t]
\centerline{\includegraphics[width=0.46\textwidth]{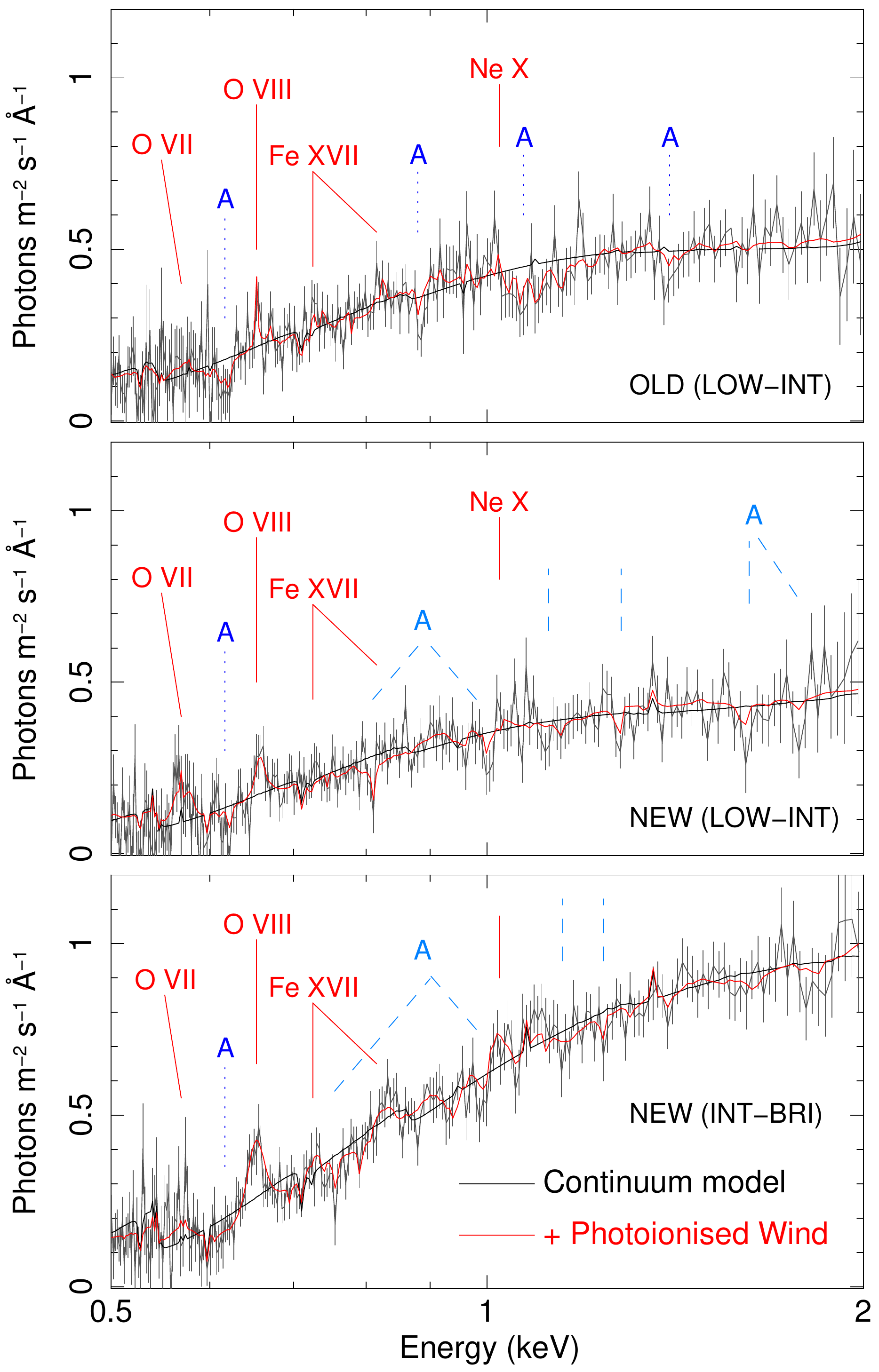}}
\caption{\;NGC 1313 ULX-1 \xmm/RGS spectra of \textcolor{black}{two fainter (HUL, top-middle)  and
              one brighter (BD, bottom) epochs} 
              with overlaid the best-fit photoionised wind  (red line) and the continuum models
              (black line, \citealt{Pinto2020b}).\label{fig:rgs}}
\vspace{-0.3cm}
\end{figure}

Spectral lines are often searched for in high-resolution spectra 
focusing on well-known He- / H-like ionic transitions.
Winds are expected to be photoionised by the source X-ray emission and,
therefore, strong lines are to be found from the dominant ionic
species such as {\oviiviii}, {\neixx} and {\nvii} in the soft X-ray band 
(0.3-2 keV) where RGS operates (see Fig.\,\ref{fig:rgs}). 

Doppler shifts from plasma motions complicate the line detection and identification.
The spectral scan with a moving gaussian line is a common procedure which indeed
identified many lines in ULX RGS spectra (see Sect.\,\ref{sec:intro}). 
However, the issue of finding spurious features, 
the so-called \textit{look-elsewhere effect}, requires thorough Monte 
Carlo (MC) simulations in order to estimate the false-alarm probability
as F-test is no longer valid \citep{Protassov2002}. 
MC method consists of simulating a large number ($\sim$10,000) of featureless continuum
spectra, which are then all searched with a gaussian line scan identical to the one used for the real data. 
This has been used extensively, providing robust significance estimates, but at a
substantial computational cost (e.g., \citealt{Pinto2020b,Kosec2018b}).

\begin{figure}[t]
\centerline{\includegraphics[width=0.48\textwidth]{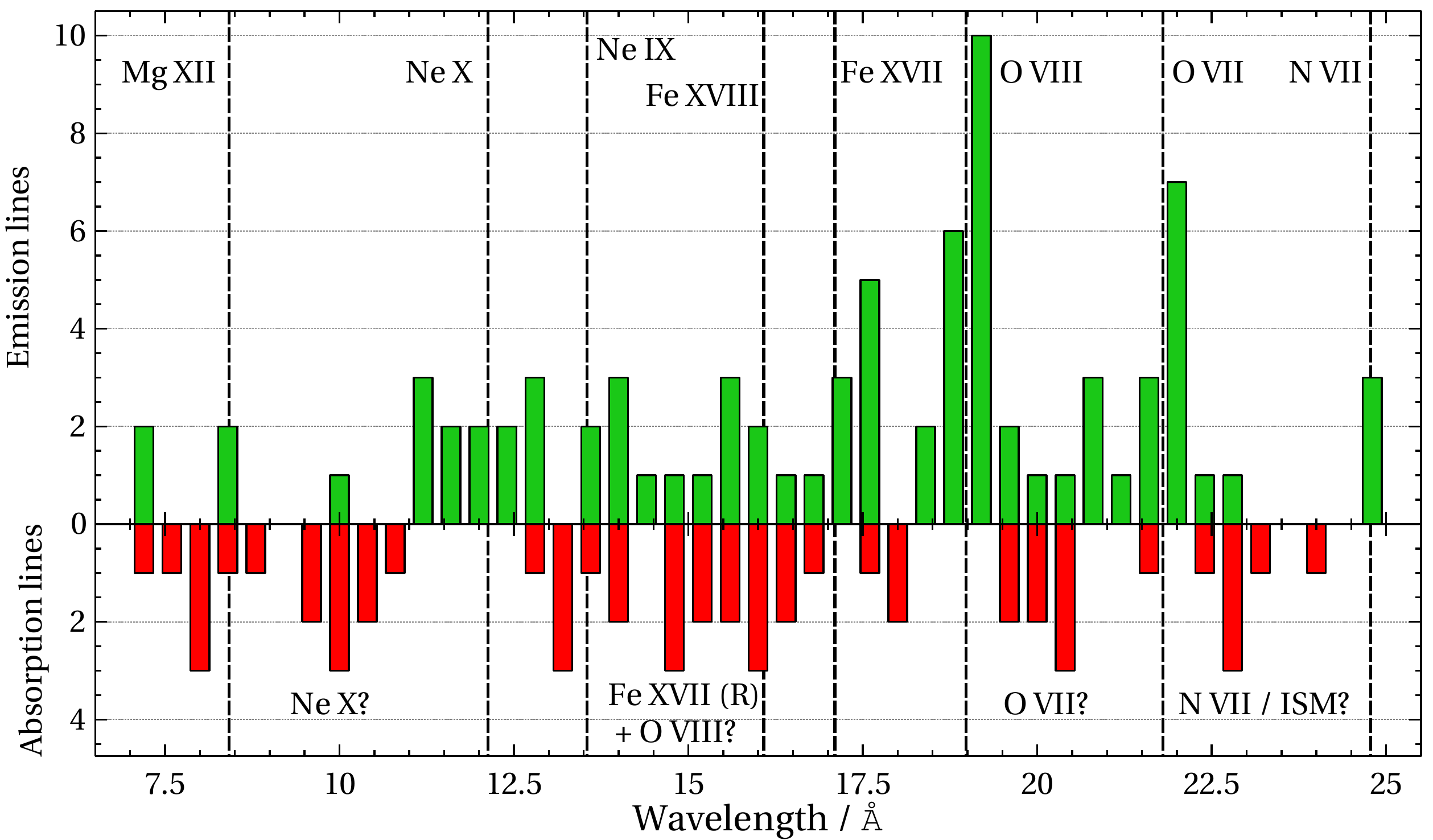}}
\caption{\;Histograms of the emission (green) and absorption (red) lines detected in a ULX sample.
Labels show the identification of frequent emission lines. The absorption lines are most
likely blueshifted (adapted from \citealt{Kosec2021}).\label{fig:sample}}
\vspace{-0.2cm}
\end{figure}

It is possible to speed up the procedure by switching from a line scan (fit) to a cross-correlation
(calculation) that compares a moving gaussian line (models) to the spectral residuals 
(both real and simulated data). 
This was successfully achieved in \citet{Kosec2021} who managed to decrease the computing time by four
orders of magnitude. 
This enabled them to perform an accurate study for a sample of 19 ULXs with high-statistics
RGS spectra and confirm the
 detection of narrow ($\lesssim1000$\,km/s) lines in the majority  of the sample
 ($>60$\%, see Fig.\,{\ref{fig:sample}}).

\subsection{Physical models: parameter-space scan}

Owing to their distance (typically above 2 Mpc) and limited fluence ($\lesssim$10$^{-12}$\,erg/s/cm$^{2}$), 
ULXs can be difficult targets for current high-spectral-resolution detectors,
due to their limited effective area. 
They are also commonly found in low-metallicity environments, which further implies weak lines.
Each one is therefore detected at low significance ($\lesssim$\,3\,$\sigma$) except for nearby ULXs 
with exposure times above 300\,ks. 
This is why lines were not detected in early observations.

Recent, more sophisticated, techniques perform a full and uniform search
of the parameter space using physical models of winds that fit multiple lines simultaneously.
The simplest case adopts gas in collisional ionisation equilibrium (CIE) and implies searching through
line of sight velocity ($v_{\rm LOS}$), velocity dispersion ($v_{\sigma}$) and electron temperature
(kT$_{\rm e}$). The normalisation or emission measure ($n_{\rm H} n_{\rm e} V$) can be a free parameter.
This makes sense when shocks between the wind and the surrounding medium or 
a thermal contribution from the jet are expected as in SS 433 \citep{Marshall2002}.
Metallicity is kept to Solar to decrease the computing time.
Besides, ULX lines are rather weak and metallicity might be unconstrained.
This was successfully applied to NGC 5204 ULX-1 and SS 433 
\citep{Kosec2018a,Pinto2021}.

\begin{figure}[t]
\centerline{\includegraphics[width=0.45\textwidth]{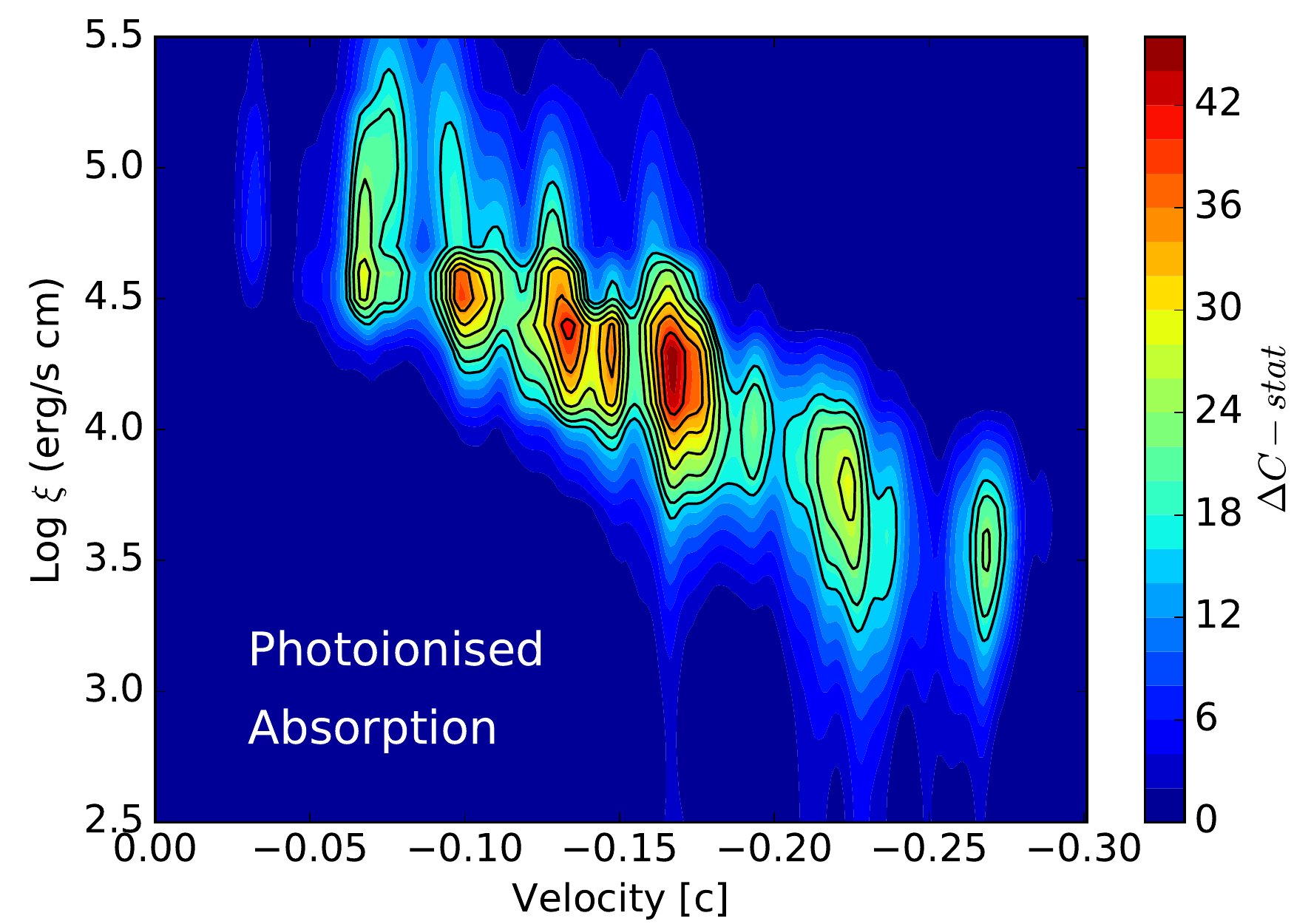}}
\vspace{-0.15cm}
   \caption{\;Photoionisation absorption model grids for NGC 247 ULX-1 {\xmm} spectrum.
                The black contours show the (2.0, 2.5, ... 5.0)\,$\sigma$ confidence levels from
                MC simulations.\label{fig:ngc247_xabs}}
\vspace{-0.3cm}
\end{figure}

Winds are most likely to be photoionised due to the strong radiation field,
at least before impacting the surrounding medium or the companion star. The variability of
the spectral features and the underlying continuum suggests that these wind 
components come from the inner disc ($10^{2-5}$\,R$_{\rm G}$, see e.g.
\citealt{Alston2021,Kara2020}). Here accurate modelling of the spectral energy distribution (SED)
is necessary in order to calculate the photoionisation equilibrium (PIE) and the ionisation parameter,
$\xi=L_{\rm ion}/nR^2$. Then $\xi$, $v_{\rm LOS}$ and $v_{\sigma}$ constitute the
backbone of the parameter space. The column density, $N_{\rm H}$, is a free parameter.
PIE scans unveiled winds in almost all ULXs with deep observations (each one lasting
$t_{\rm exp}\gtrsim$\,100\,ks or a full {\xmm} orbit, \citealt{Pinto2017,Pinto2020b};
\citealt{Kosec2018b}). In {Fig.\,\ref{fig:ngc247_xabs}} we show the remarkable example
of the detection of a photoionised absorber in the supersoft NGC 247 ULX-1 (\citealt{Pinto2021}),
blowing at 17\% of the speed of light ($c$). 
The comparison between the best-fit $\xi$ solution and the stability curves (kT or $\xi$ versus
the ratio between thermal and radiation pressure) has showed that ULX winds are likely
stable to thermal perturbations (\citealt{Pinto2020a}).

As expected, the simultaneous modelling of multiple lines by means of physically-motivated models,
significantly improves the significance, reaching peaks of 5\,$\sigma$ in the deepest observations
of ULXs with soft X-ray spectra ($\Gamma>2$).
In HUL spectral lines are significantly fainter and more difficult to detect 
(\citealt{Kosec2021}) although deep ($\gtrsim$\,100\,ks) observations 
of nearby targets have confirmed strong detections
in the HUL spectra of NGC 1313 ULX-1 and the pulsating NS NGC 300 ULX-1
(\citealt{Pinto2020b,Kosec2018b}).

The velocities of the winds (0.1-0.3\,$c$) and the ionic species ({\oviiviii} and {\neixx})
are compatible with the predictions 
from theoretical simulations \citep{Kobayashi2018}.

\section{Discussion}\label{sec:discussion}

\subsection{Super-Eddington disc-wind structure}

\begin{figure}[t]
\centerline{\includegraphics[width=0.48\textwidth]{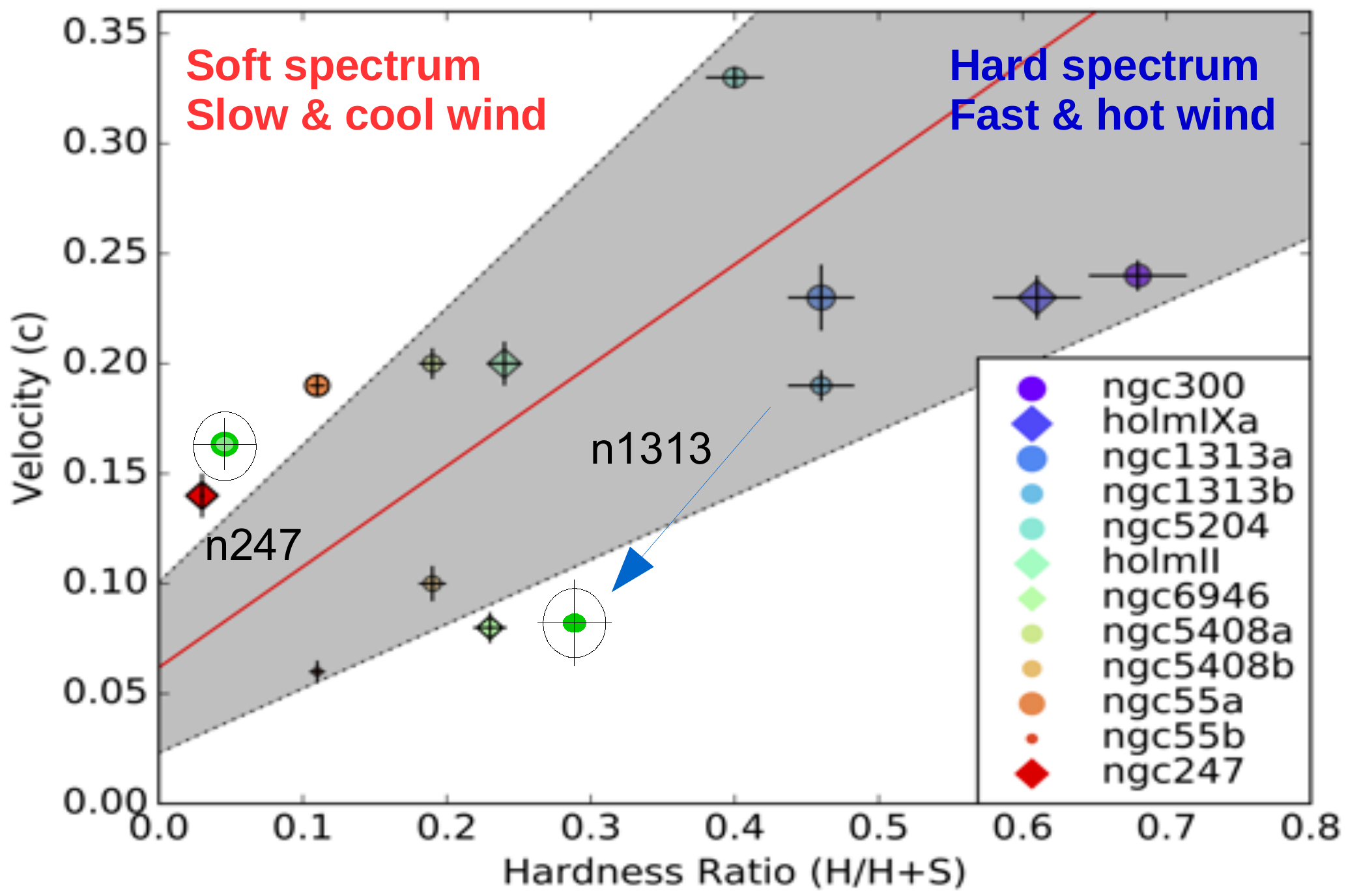}}
\vspace{-0.1cm}
   \caption{\;Correlation between LOS velocity, ionisation parameter (colour) and 
   hardness ratio (1-10 keV flux / 0.3-10 keV flux) for wind detections in different
   sources and epochs.\label{fig:trends_v_HR}}
\vspace{-0.2cm}
\end{figure}

Super-Eddington accretion predicts the \textcolor{black}{disc} to thicken and launch powerful winds
which give the system the shape of a funnel. We therefore expect at low inclinations
(i.e. face on) to see the innermost, very bright, disc region and a hard spectrum
(HUL or $\Gamma<2$). Along this LOS the wind should be highly ionised and have 
a low density. At progressively higher inclinations the wind scatters more photons out of the LOS
and the source appears softer. The outer wind portions are expected to be cooler as a consequence
of a softer SED and slower if their speed is associated to the escape velocity 
$v_{\rm esc}=\sqrt{2GM/R}$. This was confirmed, albeit with a limited source sample, by
\citet{Pinto2020a} in the form of a positive correlation between the wind velocity, the 
ionisation parameter and the ULX spectral hardness (see {Fig.\,\ref{fig:trends_v_HR}}).
This is also validated by the spectra taken at different epochs for a few ULXs.

The scenario is likely more complex as the opening angle of the funnel depends
on the mass accretion rate. Thus, the observed wind properties depend on a combination 
of the inclination angle and the Eddington ratio.

It is important to notice that optical spectroscopy reveals the presence of strong lines 
from H- and He- transitions in many ULXs. The lines are very similar to those observed 
in SS 433 and have been associated with the cool, outer, photosphere of the super-Eddington disc wind
($10^{5-6}$\,R$_{\rm G}$, see e.g. \citealt{Vinokurov2018, Fabrika2021})
and further confirm that SS 433 is most likely a highly obscured ULX 
that is being seen almost edge-on as shown by Fe K time lags \citep{Middleton2021}.

\subsection{Feedback and growth rate}

Owing to their extreme velocities, ULX winds are expected to have a significant impact
on their environments. Many ULXs, including NGC 1313 ULX-1,
are indeed surrounded by huge interstellar cavities or bubbles
with supersonic expansion \citep[80-250 km/s,][]{Gurpide2022}. 
The bubbles are young ($10^{5-6}$ yr) and their H\,$\alpha$ luminosities 
require a mechanical power of $10^{39-40}$\,erg/s \citep{Pakull2002}.

The kinetic power of the winds can be written as follows: 
$L_w = 0.5 \, \dot{M}_w \, v_w^2 =  2 \, \pi \, m_p \, \mu \, \Omega \, C \, L_{\rm ion} \, v_w^3 \, / \, \xi$
\textcolor{black}{where $\dot{M}_w = 4 \, \pi \, R^2 \, \rho \, v_w \, \Omega \, C$ is the outflow rate,} $\Omega$ 
and $C$ are the solid angle and the volume filling factor (or \textit{clumpiness}), respectively,
$\rho=n_{\rm H}  \, m_p \, \mu$ is the density and $R$ is the distance from the ionising source.
The fraction of ULXs with wind detections is $>60\%$ (\citealt{Kosec2021}) which implies
a solid angle $\Omega/4\pi\gtrsim0.3$ in agreement with theoretical simulations. 
The typical ionisation parameters found in ULX winds (log $\xi\sim$ 2-4) would correspond to 
a clumping factor between 0.01-0.1 \citep{Kobayashi2018}.
This yields a kinetic power $L_w = 10^{39-40} \, {\rm erg/s}$,
which is likely enough to inflate the superbubbles.

ULX winds also take away of lot of momentum and matter from the system.
Comparing their kinetic power with the bolometric luminosity we estimate that about 
50\% of the energetic budget is lost to launch the winds. 
The maximum outflow rate $\dot{M}_w$ is $\sim$\,90\% (assuming standard accretion efficiency,
i.e. $\eta=0.1$). However accounting for advection and photon trapping we estimate a more 
realistic mass loss rate of 10-50\% \citep{Mushtukov2019a}.
The compact object would still grow fast but at a milder super-Eddington rate.

\subsection{Current limitations and prospects}

There are some limitations in the current facilities that hamper a full understanding of the 
ULX phenomenology. We know that spectral lines vary over the time and in response to the
variability of the underlying continuum, showing higher fluxes and significance in softer spectra
(\citealt{Pinto2020b,Kosec2021}).
Accurate physical modelling has shown that the wind is cooler and slower in softer spectra perhaps
due to a larger launching radius at higher accretion rates (\citealt{Pinto2021}). 
The picture is more complex as the emission lines become stronger and require multiple components
likely from different regions (\citealt{Pinto2020b}).
\textcolor{black}{It is difficult to} distinguish among different processes (variable accretion rate, precession, 
stochastic variability, etc.) because current facilities require to integrate well over 100 ks of exposure
thereby preventing us from resolving the variability timescales which can be as short as 
$100$\,s (e.g. \citealt{Alston2021,Kara2020,Kobayashi2018}).
Another issue is the lack of sensitivity and resolution in the Fe K band (6-9 keV) which limits
our capability of detecting hot wind components from the inner disc expected
for ULXs with hard spectra (HUL). Presently, the only Fe K detections
are those for NGC 1313 ULX-1, NGC 300 ULX-1 and NGC 4045 HLX-1
\citep{Walton2016a, Kosec2018b, Brightman2022}.
\textcolor{black}{Of course, deep observations of other ULXs would help to increase the Fe K line sample.}

\begin{figure}[t]
\centerline{\includegraphics[width=0.46\textwidth]{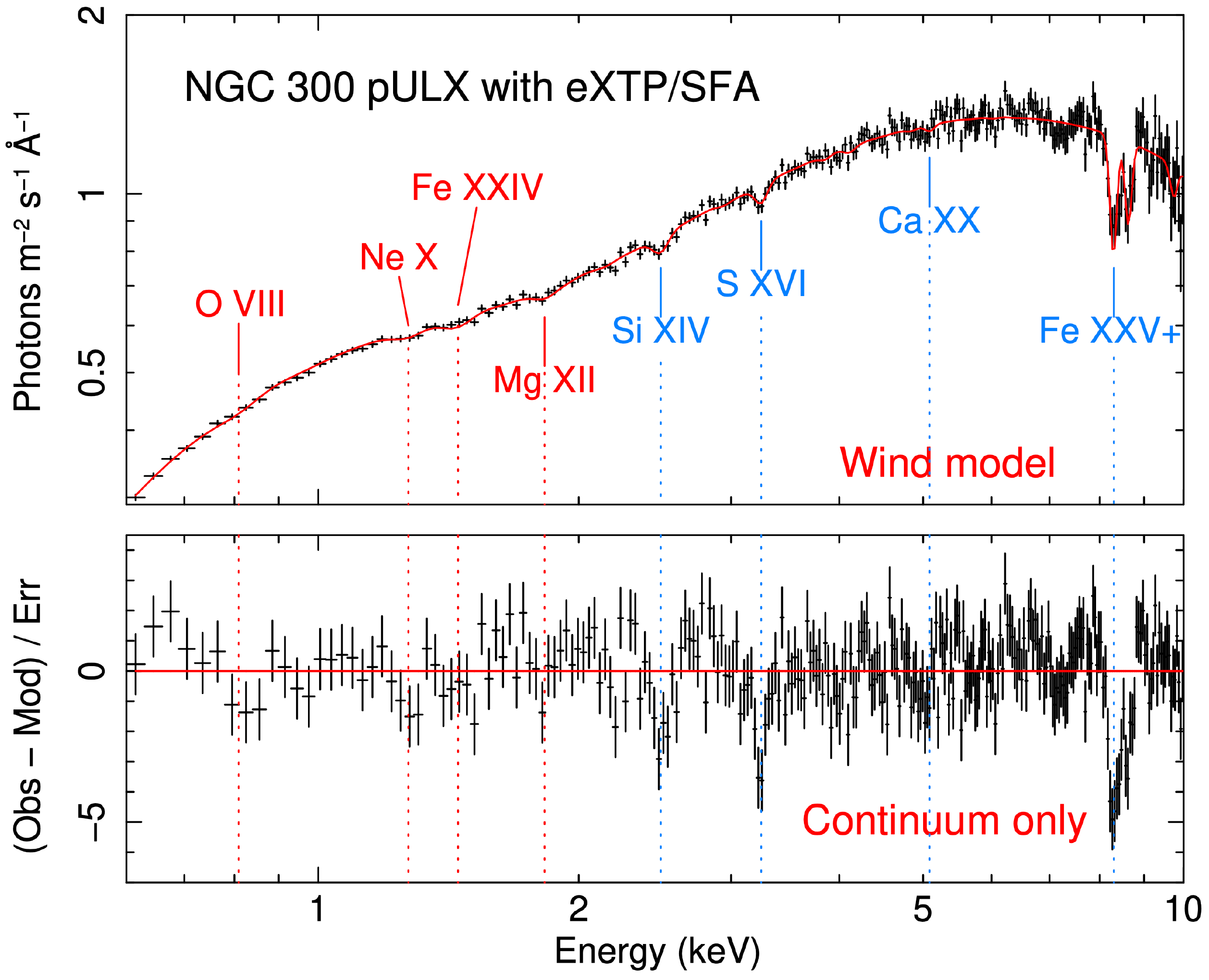}}
   \caption{\;Pulsating NGC 300 ULX-1 eXTP/SFA simulation (80\,ks, top panel) 
                adopting the best-fit continuum
                 plus wind model from \citet{Kosec2018b}.
                 \textcolor{black}{For the residuals plot (bottom panel)
                  the wind component was removed from the model.} \label{fig:eXTPsim}}
\vspace{-0.3cm}
\end{figure}

The enhanced X-ray Timing and Polarimetry ({\extp}) mission (2027-, \citealt{Zhang2019}) will 
provide a high effective area in the hard X-ray band ($>$\,2\,keV) which together with a low 
background due to the low Earth orbit \textcolor{black}{could significantly} improve the detection of the 
Fe K wind component with respect to {\xmm}. This is showcased in Fig.\,\ref{fig:eXTPsim} with \textcolor{black}{our} 80\,ks simulation of NGC 300 ULX-1 using the Spectroscopic Focusing Array (SFA).
\textcolor{black}{We estimate that the} {\fexxv} absorption line may be detected at $5\,\sigma$. 
Physical models gathering multiple lines will achieve this at much shorter exposure times.

\begin{figure}[t]
\centerline{\includegraphics[width=0.46\textwidth]{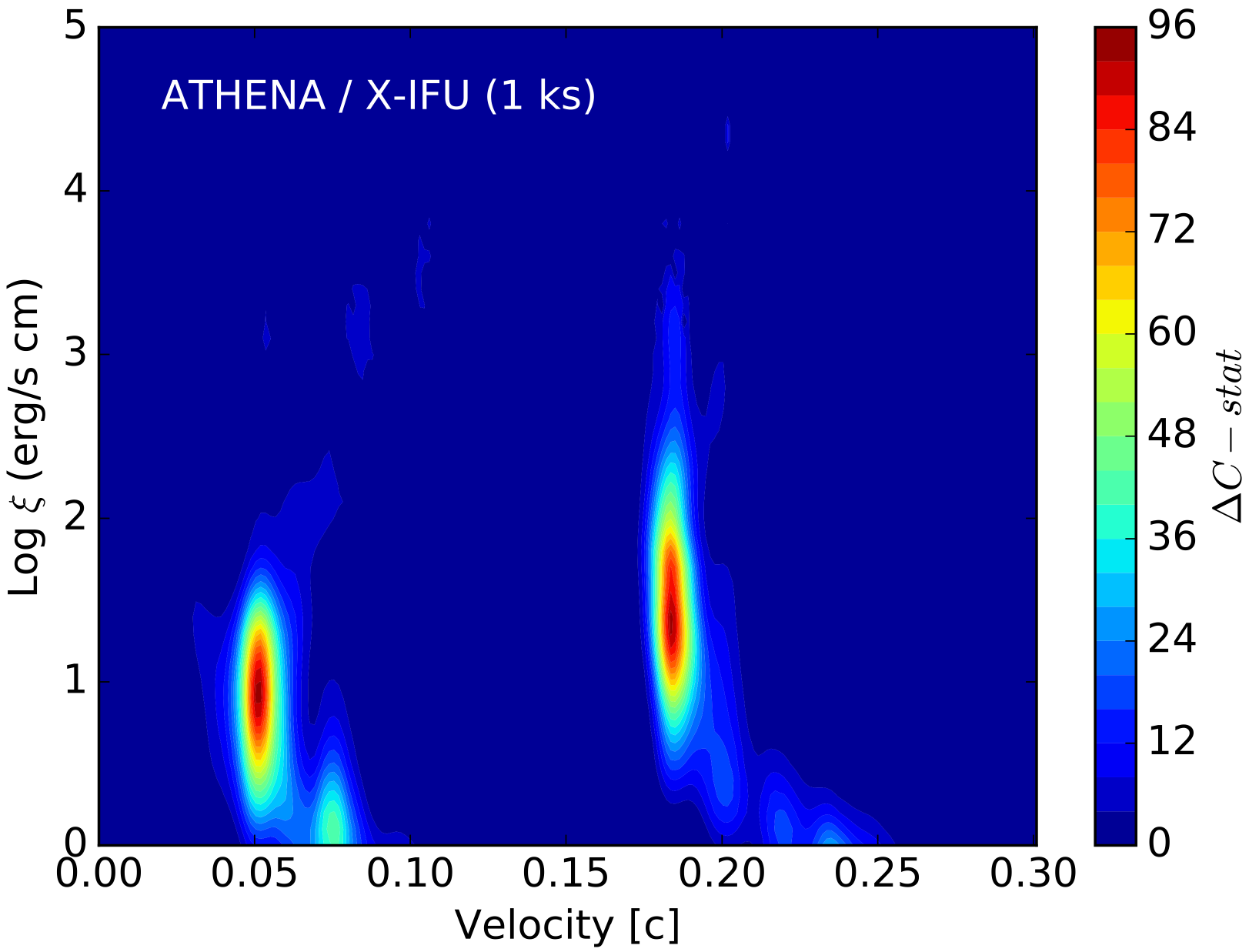}}
\centerline{\includegraphics[width=0.46\textwidth]{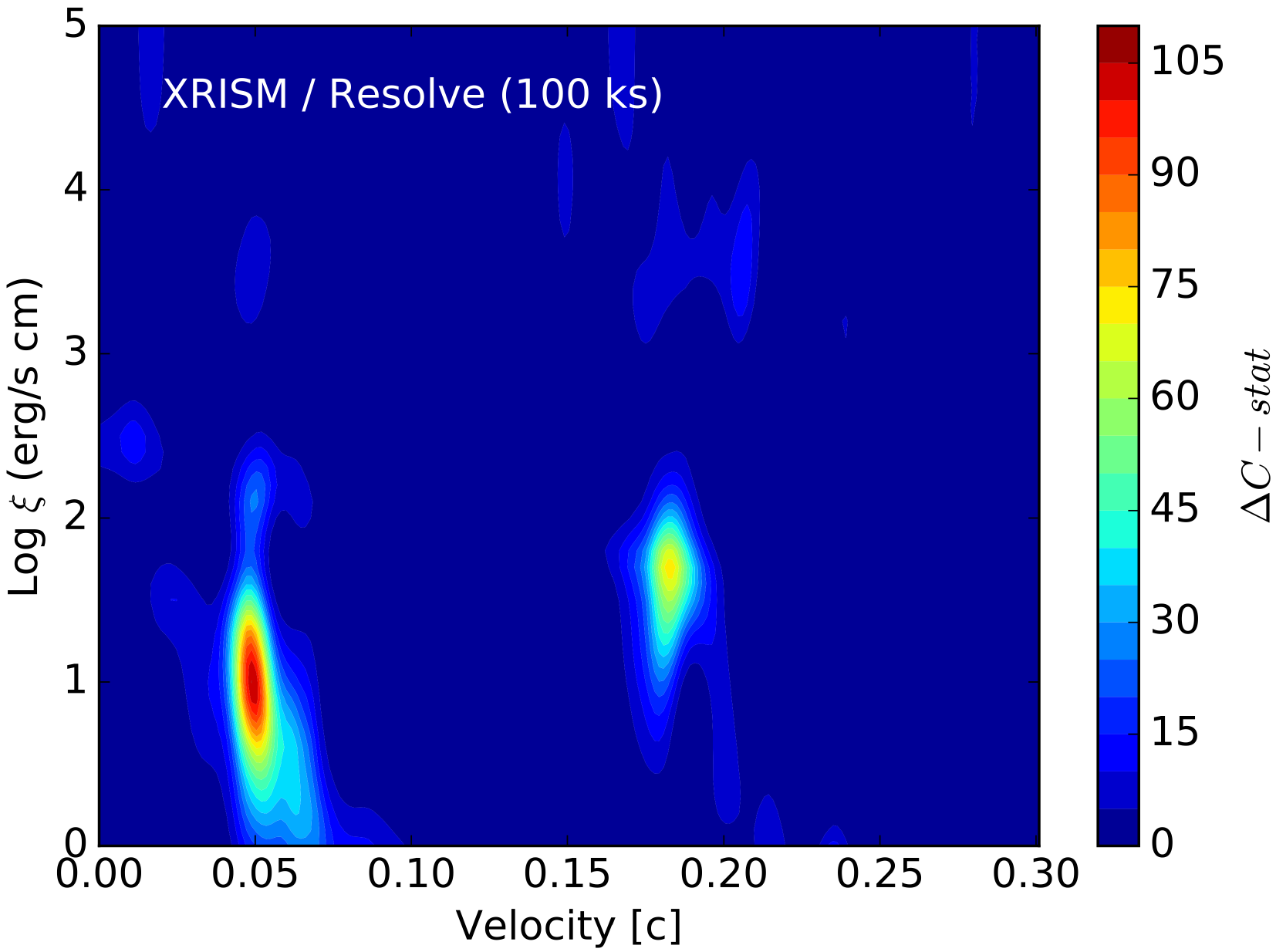}}
   \caption{\;NGC 1313 ULX-1 simulations with Athena/X-IFU and XRISM/Resolve microcalorimeters.
                 Adopted is the best-fit continuum plus wind model from \citet{Pinto2020b}
                 for the bright regime (see also Fig.\,\ref{fig:rgs} bottom panel).\label{fig:microsim}}
\vspace{-0.3cm}
\end{figure}

Microcalorimeter spectrometers will boost our sensitivity to weak lines. This can be defined as 
the product between \textcolor{black}{effective area and resolving power} or $A_{\rm eff} / \Delta E$.
The X-Ray Imaging and Spectroscopy Mission ({\xrism}, 2023-, \citealt{Guainazzi2018})
will bear the Resolve microcalorimeter ($\Delta E \sim5$eV and PSF $\sim1'$) 
which, similarly to \textit{Hitomi}/SXS, is
10 times more sensitive than the gratings aboard {\xmm} and {\chandra}
above 4 keV, and even more above 7 keV.
In Fig.\,\ref{fig:microsim} (bottom panel) we show a physical model scan of a 100\,ks 
{\xrism}/Resolve spectrum \textcolor{black}{that we have} simulated for NGC 1313 ULX-1 using the best-fit continuum model 
plus multiphase photoionised wind.
Both cold/slow and warm/fast components are detected at very high significance. 
For the brightest, nearby, ULXs exposures of a few 10\,ks should be sufficient to
achieve $3\sigma$ detections and enable studies of line variability within half a day,
which is currently impossible.

A groundbreaking improvement \textcolor{black}{would} be provided by the X-ray Integral Field Unit (X-IFU)
aboard the Advanced Telescope for High-ENergy Astrophysics ({\athena}, 2035-, \citealt{Barret2022}).
\textcolor{black}{At the time of writing, X-IFU is designed to achieve an} exquisite combination of high effective area (1m$^{2}$) and spectral resolution ($\lesssim2.5$eV),
small PSF (5-10"), and low background. \textcolor{black}{This would} 
provide a 2-orders-of-magnitude improvement 
with respect to the current instruments, about 10 times better than {\xrism}/Resolve 
or above given that ULXs are often found in crowded areas.
In Fig.\,\ref{fig:microsim} (top panel) we show a similar simulation for the bright state
of  NGC 1313 ULX-1 using {\athena}/X-IFU and adopting an exposure time of just 1\,ks.
This snapshot would be sufficient to detect the wind components at very high significance 
\textit{de-facto} enabling, for the first time, the simultaneous study of the variability in the wind
and the underlying spectral continuum. 
\textcolor{black}{This can be extended to other ULXs and}
is the key to answer the main question regarding the nature of the ULX regime transitions
(SUL $\leftrightarrow$ BD $\leftrightarrow$ HUL), the role of the wind, its location and 
launching radius.

The codes will have to be adapted to the size and complexity of the response matrices 
of future detectors and parameter space. The use of cross-correlation and machine learning may 
provide a viable solution. Recently, it was also shown that the slope of the histogram of the 
$\Delta C$-stat or $\Delta \chi^2$ remains constant above a certain number of simulations 
(\citealt{Pinto2021}), which \textcolor{black}{may} therefore 
be extrapolated and used to estimate higher values of significance at low computational costs.

\section{Conclusions}\label{sec:conclusions}

We have provided a brief overview on the properties of winds in ULXs.
As the most valuable candidates of super-Eddington accreting systems, ULXs
are predicted to blow powerful winds that slow down the accretion onto the compact object
and affect the surrounding medium. 
Current instruments, primarily {\xmm}/RGS, enabled to obtain highly significant detections
and a qualitative view of the wind structure and general properties.
An intimate connection between the wind and the source state is suggested, potentially
unveiling the detailed accretion and variability mechanism. However,
a larger statistical sample of ULX winds and more observations per target are
required to place more quantitative constraints.  
Our understanding is also limited by the variability timescales that are 
too short to be sampled with the current instruments.
Future missions such as \textit{XRISM}, \textit{eXTP} and {\athena} will \textcolor{black}{
boost our sensitivity to such timescales and to different thermal phases of the wind,
providing a better coverage of its parameter space.} 


\section*{Acknowledgments}

We acknowledge support from the European Union’s Horizon 2020 Programme under the 
AHEAD2020 project (grant agreement n. 871158), ESA Research Proposals programme
and NASA grants 80NSSC21K0872 and DD0-21125X.

\bibliography{Wiley-ASNA}%

\end{document}